# Superconductivity in electron doped PbBi$_2$Te$_4$

Xianghan Xu[1,*], Danrui Ni[1], Weiwei Xie[2], and R.J. Cava[1,*]

1. Department of Chemistry, Princeton University, Princeton, NJ 08648
2. Department of Chemistry, Michigan State University, Lansing, MI 48824

**Abstract**

Single crystals of In-doped PbBi$_2$Te$_4$ are synthesized via a conventional solid-state method. Chemical analysis and hall measurements indicate that In replaces Pb, introducing *n*-type carriers, creating Pb$_{1-x}$In$_x$Bi$_2$Te$_4$. A superconducting transition is observed with a maximum transition temperature around 2.06 K for Pb$_{1-x}$In$_x$Bi$_2$Te$_4$. Field dependent transport measurements reveal type-II superconductivity and yield a maximum upper critical field around 1.55 T. Thermodynamic data indicates bulk superconductivity in the BCS weak coupling limit. Our findings establish an ambient-pressure superconducting system in the AM$_2$X$_4$ family, and doped PbBi$_2$Te$_4$ as a promising platform for the study of topological superconductivity.

**Introduction**

The emergence of superconductivity in topological materials is predicted to open an energy gap in the surface states and lead to the exotic Majorana zero modes[1-3], with accompanying potential applications in topological quantum computing[4,5]. Although *p*-wave superconductivity is believed to be ideally compatible with a topological surface state[6], it has also been demonstrated that the proximity effect enables the appropriate interactions between superconductivity and topological states for an *s*-wave superconductor at the interfaces of heterojunctions[2,7], as well as at the geometrical surfaces of single-phase materials[8-11]. Therefore, the exploration of new superconductors with topological features in their energy bands is prominent in condensed matter physics research.

Tuning the Fermi level by pressure or chemical doping is a feasible way to induce superconductivity in topological materials. For example, Bi$_2$Se$_3$ and Bi$_2$Te$_3$ are well studied topological insulators[12-14], and bulk superconductivity can be induced in them via applying high pressure[15,16], or introducing intercalated dopants such as Cu[17], Pd[18], Nb[19], and Sr[20]. The related AM$_2$X$_4$ (A=Mn, Ge, Sn, Pb; M=As, Sb, Bi; and X=Se, Te) family has a centrosymmetric space group *R-3m* and a crystal structure derived by an alternate stacking of AX and M$_2$X$_3$ blocks, adding an additional structural degree of freedom on top of the topology. For example, MnBi$_2$Te$_4$ and related materials successfully combine intrinsic magnetism and topology in single-phase bulk materials [21,22]. Recently, researchers reported bulk superconductivity in a series of In-doped SnBi$_2$Te$_4$ crystals (maximum $T_c$ ~ 1.85 K) [23], which sheds light on a new family of topological superconductor candidates. Thus, synthesis and characterizations of more superconductors in the AM$_2$X$_4$ family is urgently desired. As another member in this family, PbBi$_2$Te$_4$ has been proved experimentally to be a 3D topological insulator [24] and pressure-induced superconductivity has been observed in PbBi$_2$Te$_4$[25]. Indium is known as a good aliovalent dopant when partially replacing Pb[26]. For example, high-pressure-synthesized Pb$_{0.2}$In$_{0.8}$Te cubic phase is reported to be superconducting below $T_c$ ~ 4.8 K[27]. Therefore, the potential emergence of superconductivity in Pb$_{1-x}$In$_x$Bi$_2$Te$_4$ is a promising possibility.

Here, we report the synthesis of Pb$_{1-x}$In$_x$Bi$_2$Te$_4$ (*x* = 0, 0.4, 0.5, 0.6, 0.7) crystals using a conventional solid-state reaction method. Superconducting transitions are observed from *x* = 0.4 to

$x = 0.7$. The superconducting $T_c$ increases monotonically with increasing In doping concentration, and reaches maximum $T_c \sim 2.06$ K for $x = 0.7$. Thermodynamic measurements suggest bulk superconductivity as well as weak coupling BCS nature for the superconducting state. Hall measurements suggest that the parent compound $PbBi_2Te_4$ is *n*-type, and that the In replacement of Pb introduces more electrons into the system, which is opposite to the case reported for In-doped $SnBi_2Te_4$[23]. Our results establish a new topological superconductor candidate in the $AM_2X_4$ family. The $Pb_{1-x}In_xBi_2Te_4$ single crystals also provide excellent opportunities for future spectroscopic studies to investigate the interplay between superconductivity and topology.

**Experimental**

*Crystal growth*: Stoichiometric amounts of Pb, In, Bi, and Te were weighed and placed in an alumina crucible. The crucible was sealed in a vacuum quartz tube, heated to 600°C, held for 100 hours, and then furnace-cooled to room temperature. The obtained ingot was then ground, poured into an alumina crucible, and sealed in a vacuum quartz tube. The tube was heated to 590°C, kept for 50 hours, then cooled to 300°C at 2°C/h. The crystal ingot was mechanically separated from the crucible. Al and Si were not found to be present in the crystals in the EDS characterization, suggesting negligible contamination from the crucible materials.

*Electric, magnetic, and thermodynamic measurements*: The data were collected in a Dynacool PPMS-9. Electrode contacts were made by silver epoxy. Measurements below 1.8 K used a He-3 refrigerator.

*Structural and chemical characterizations*: Powder x-ray diffraction was carried out at room temperature using a Bruker D8 Advance Eco diffractometer with a Cu anode. Single-crystal x-ray diffraction was carried out using a Bruker D8 Quest Eco diffractometer equipped with a Photon III CPAD detector and monochromated Mo K$\alpha$ radiation ($\lambda = 0.71073$ Å). The frames were integrated using the SAINT program within the APEX III 2017.3-0 operating system. The structure was determined using direct methods and difference Fourier synthesis (SHELXTL version 6.14) [28]. The reciprocal lattice planes from single-crystal x-ray diffractions are displayed in Fig. S4, and the $|F_{cal}|^2$ vs. $|F_{obs}|^2$ plots from the refinement results are shown in Fig. S5. Scanning electron microscopy (SEM) images and Energy-dispersive X-ray spectroscopy (EDS) were collected using a Quanta 200 FEG Environmental-SEM on the cleaved surfaces of the $Pb_{1-x}In_xBi_2Te_4$ crystals.

**Results and discussion**
**i. crystal structure**

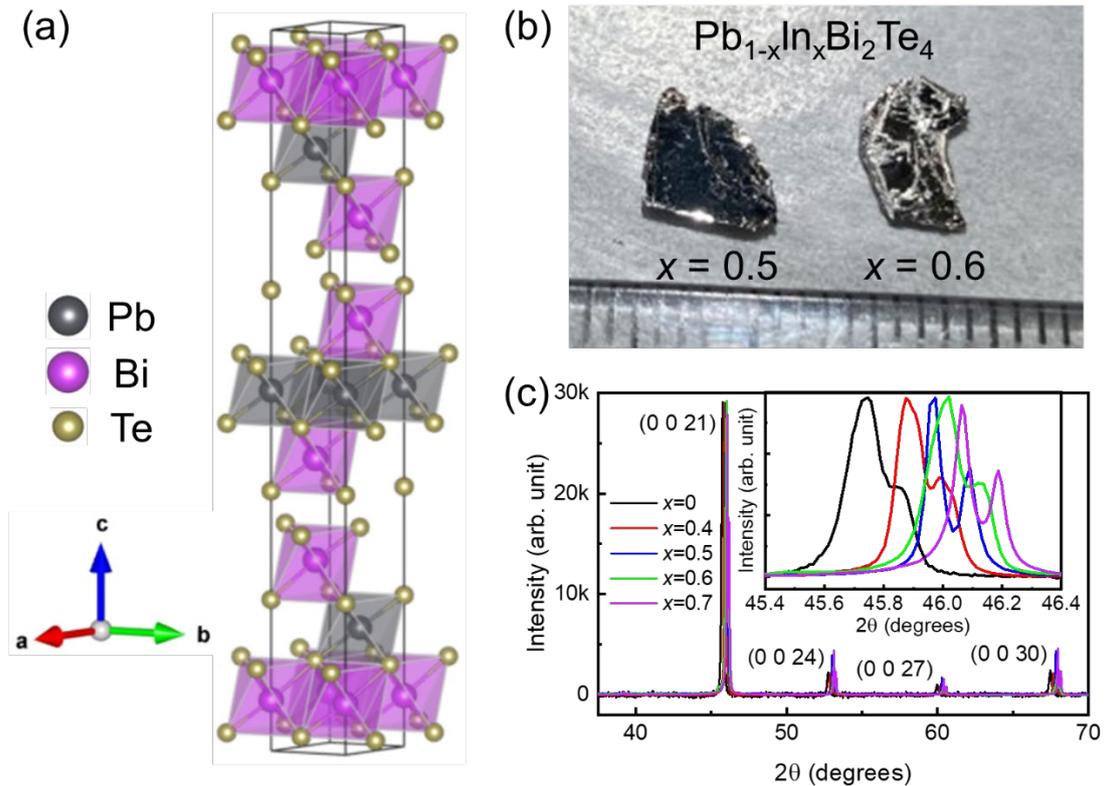

**Figure 1.** (a) The idealized crystal structure of the parent phase $PbBi_2Te_4$. (b) Photos of $Pb_{1-x}In_xBi_2Te_4$ crystals with cleaved *ab* planes. (c) X-ray diffraction of (0 0 L) peaks on cleaved ab planes. The inset shows the systematic shift of the (0 0 21) peak position to a higher $2\theta$ angle with increasing In doping concentration.

As shown in Table 1, the Pb: In: Bi: Te ratios determined from the EDS experiments are within experimental error of the nominal compositions, suggesting that the nominal stoichiometry of the crystals studied in the present work is correct. The SEM images and EDS spectra are shown in Fig. S1[29]. Fig. 1a displays the crystal structure of the parent phase $PbBi_2Te_4$ ($x = 0$) solved by single crystal X-ray diffraction (SC-XRD). The structure can be regarded as an alternate stacking of a [PbTe] layer and a [$Bi_2Te_3$] layer, in where the Pb and Bi atoms are located in Te octahedra. It is impossible to distinguish Pb and Bi unambiguously by X-ray diffraction due to their similar electron counts, and the refinement assumed a fully ordered Pb/Bi distribution, which has been suggested in previous studies[25,30]. Fig. 1b displays photographs of several of the $Pb_{1-x}In_xBi_2Te_4$ crystals. They are easy to cleave due to the layered structure. As shown in Fig. 1c, the position of the (0 0 L) diffraction peaks monotonically shift to a higher angle as the In doping concentration increases, which supports the EDS results and indicates that the In atoms are incorporated into the lattice and cause the lattice to shrink. Although Pb and Bi are indistinguishable by X-ray diffraction, the crystallographic sites for the In (having smaller atomic number) are detectable. The SC-XRD refinement results on In occupancies are shown in Table S1, which indicate that the In atoms go to both Pb sites (Wyckoff position 3*a*) and Bi sites (Wyckoff position 6*c*) with occupancies in the same order of magnitude. The In concentrations derived from the SC-XRD refinements are also close to the nominal compositions, which confirm the reliability of the refined In occupancies. Based on this fact, even if the Pb and Bi fully order in the parent phase $PbBi_2Te_4$, introducing In will inevitably cause Pb-Bi mixing, and that type of disorder does not change the stoichiometry. Note that the $x = 0.7$ crystal starts to show tiny impurity peaks in the XRD pattern (Fig. S2). The heights of those impurity peaks are less than 1% of the real peaks, which implies that the impurities are present in tiny amounts and should not influence the quality of $x = 0.7$ crystal studied here. But it does mean that 70% is close to the upper limit of In solubility in $PbBi_2Te_4$ when the material is prepared by conventional solid-state synthesis. The attempt to synthesize $InBi_2Te_4$ using a similar conventional solid-state reaction method resulted in a mixture with $Bi_2Te_3$ and $In_2Te_3$ as major phases.

**Table 1**. Experimental parameters of $Pb_{1-x}In_xBi_2Te_4$ crystals.

| Nominal composition | EDS composition | $a$ (Å) | $c$ (Å) | $n$ (cm$^{-3}$) (all $n$-type) | $T_c$ (K) | $H_{c2}(0)$ (T) |
|---|---|---|---|---|---|---|
| $PbBi_2Te_4$ | $PbBi_{1.97}Te_{3.81}$ | 4.4360(10) | 41.636(7) | 4.23e+20 | -- | -- |
| $Pb_{0.6}In_{0.4}Bi_2Te_4$ | $Pb_{0.61}In_{0.39}Bi_{2.26}Te_{4.32}$ | 4.4172(4) | 41.498(1) | 4.99e+20 | 1.31(8) | 0.87(8) |
| $Pb_{0.5}In_{0.5}Bi_2Te_4$ | $Pb_{0.50}In_{0.50}Bi_{2.15}Te_{4.15}$ | 4.4143(5) | 41.425(8) | 2.25e+21 | 1.53(1) | 0.99(4) |
| $Pb_{0.4}In_{0.6}Bi_2Te_4$ | $Pb_{0.41}In_{0.59}Bi_{1.99}Te_{3.72}$ | 4.4177(4) | 41.405(3) | 3.46e+21 | 1.91(6) | 1.47(1) |
| $Pb_{0.3}In_{0.7}Bi_2Te_4$ | $Pb_{0.33}In_{0.67}Bi_{1.80}Te_{3.73}$ | 4.4067(9) | 41.344(9) | 3.47e+21 | 2.06(7) | 1.55(2) |

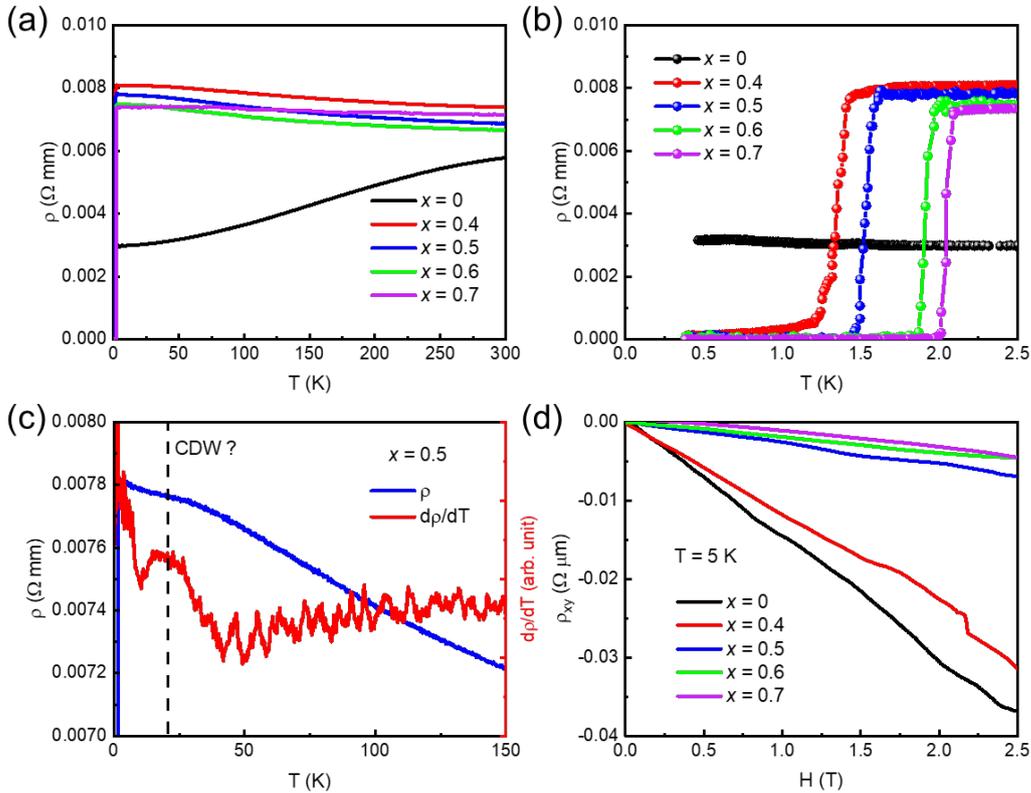

**Figure 2.** (a) Resistivity as a function of temperature curves. (b) Low temperature regime of (a). (c) The resistivity and its derivative of $Pb_{0.5}In_{0.5}Bi_2Te_4$. (d) Hall resistivity as a function of magnetic field measured at 5 K (normal state).

**ii. electrical properties and superconductivity**

The resistivity vs. temperature ($\rho$-$T$) curves down to around 0.4 K of $Pb_{1-x}In_xBi_2Te_4$ crystals are displayed in Fig. 2a. The metallic behavior and resistivity value of the $x = 0$ crystal studied here are consistent with the previous report. In the In-doped crystals, the $\rho$-$T$ data show weakly semiconducting behavior. As shown in Fig. 2b, a sharp superconducting transition is detected in the $x = 0.4$ to 0.7 crystals. The transition temperature ($T_c$, defined by the temperature where the resistivity drops 50%) monotonically increases with increasing In concentration, reaching a maximum of ~ 2.06 K for the $x = 0.7$ crystal. The DC magnetic susceptibility data for $x = 0.7$ (Fig. S3) shows a drop below $T_c$, and the divergence between zero-field-cool and field-cool curves suggests Type-II superconductivity. Note that the susceptibility does not saturate down to around 1.7 K, which is the lowest temperature that our VSM can reach. As shown in Fig. 2c, the $\rho$-$T$ of $x = 0.5$ crystal has an anomaly at around 20 K, which we tentatively attribute to a change density wave (CDW) transition. Interestingly, this feature was only observed in the Pb: In = 1: 1 sample, but was not found in samples with other In concentrations. Fig. 2d shows the result of Hall measurements at 5 K. The negative Hall coefficients suggest all $Pb_{1-x}In_xBi_2Te_4$ crystals are $n$-type, and that a larger fraction of In doping introduces more electrons, indicated by the decreasing absolute value of the Hall coefficient. The carrier densities calculated from Hall coefficients are displayed in Table 1. Undoped $PbBi_2Te_4$ has been reported to be $n$-type[25], which is consistent with the present work. Commonly, the valence of In can be either +1 or +3. Previous studies suggest that In is mainly +1 in In-doped SnTe and introduces $p$-type carriers[31], but the In is mainly 3+ in In-doped PbTe and introduces $n$-type carriers[32,33]. Similarly, In-doped $SnBi_2Te_4$ is reported to be $p$-type[23], an In-doped $PbBi_2Te_4$ in the present work is found to be $n$-type. One difference is that In-doped SnTe is superconducting[31], but In-doped PbTe is not superconducting in the Pb-rich regime[26], whereas In-doped $SnBi_2Te_4$ and $PbBi_2Te_4$ all show superconducting transitions at similar temperatures. Note that, although the Hall results confirm dominant $n$-type conduction in $Pb_{1-x}In_xBi_2Te_4$, the non-linearity of the carrier density versus doping concentration suggests the presence of degenerate In valence states (mixture of +1 and +3) or additional impurities which act in a self-compensating manner [34]. Fig. 3 shows the $\rho$-$T$ curves measured in various magnetic fields parallel to the $c$ axis. The superconducting transitions are getting less sharp with increasing field in all samples, as is typical for type-II superconductivity. Fig. 4 summarizes the field dependence of the superconducting transition temperatures obtained from the $\rho$-$T$ curves The upper critical field $H_{c2}$ (0) is determined by fitting the data points using the following equation based on generalized Ginzburg-Landau model:

$H_{c2}(T) = H_{c2}(0)[(1-t^2)/(1+t^2)]$, where $t = T/T_c$. The fitting results are displayed in Table 1. The obtained $H_{c2}$ (0) is 1.55(2) T in the highest $T_c$ ($x = 0.7$) sample, which is close to the reported $H_{c2}$ (0) of $Sn_{0.39}In_{0.61}Bi_2Te_4$[23]. Then, the Ginzburg−Landau coherence length, $\xi_{GL}$, is calculated to be 193 Å, 182 Å, 150 Å, and 145 Å for $x = 0.4, 0.5, 0.6$, and 0.7, respectively, based on Ginzburg-Landau relation: $H_{c2}(0) = \dfrac{\Phi_0}{2\pi\xi_{GL}^2}$, where $\Phi_0 = hc/2e$ is the flux quantum.

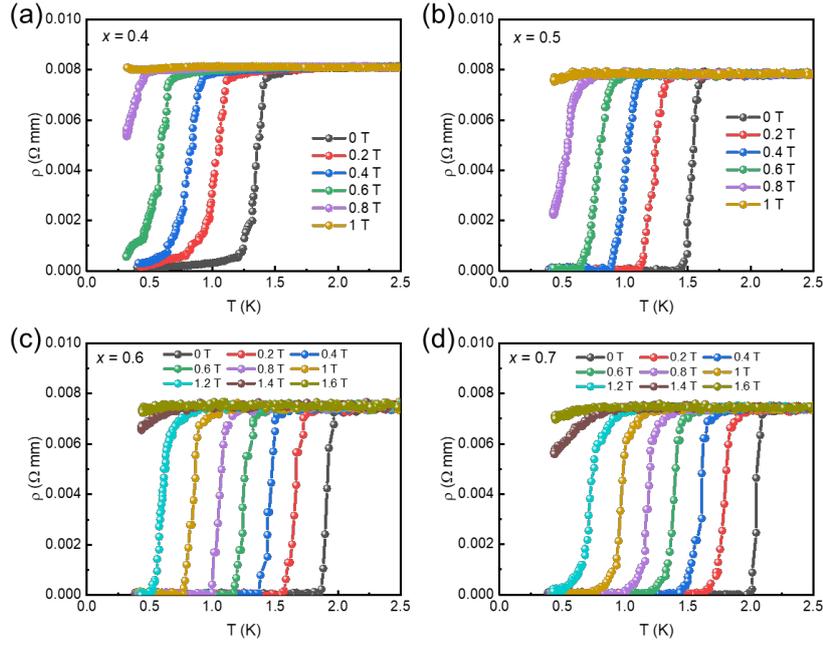

**Figure 3.** Resistivity as a function of temperature for single crystals of $Pb_{1-x}In_xBi_2Te_4$ in various magnetic fields. (a) $x$ =0.4 (b) $x$ =0.5 (c) $x$ =0.6 (d) $x$ =0.7.

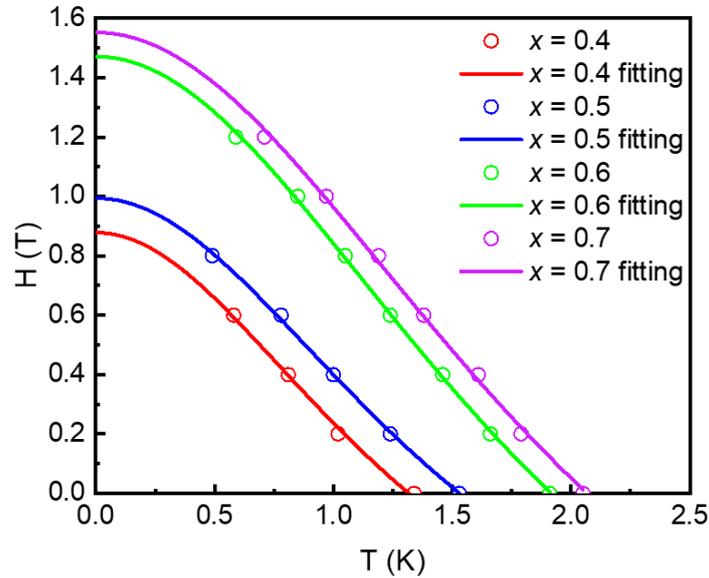

**Figure 4.** Temperature dependence of the upper critical field $H_{c2}$ (open circles) and the upper critical field fitting (solid lines).

### iii. Thermodynamics

Fig. 5a shows the zero-field heat capacity as a function of temperature for $x$ = 0.5, 0.6, and 0.7 down to around 0.5 K. Clear anomalies can be seen at the superconducting transitions. Normal state heat capacity data (from 2.3 K to 5 K) are fitted using $C_p = \gamma T + \beta T^3 + \delta T^5$, which yields the Sommerfeld coefficients $\gamma$ = 3.5(6), 4.1(7), and 4.2(1) mJ/mole/K$^2$ for $x$ = 0.5, 0.6, and 0.7, respectively. The γ values for In-doped PbBi$_2$Te$_4$ are slightly smaller than the reported values in superconducting In-doped SnBi$_2$Te$_4$[23]. Note that the normal state heat capacity cannot be fitted well without the $T^5$ term, which may suggest the existence of anharmonic phonons in this system. (In fact, the $T^5$ term is also claimed to be necessary to explain the normal state heat capacity of In-doped SnBi$_2$Te$_4$[23].) Fig. 5b displays $C_p$/T after subtracting the normal electron and phonon contributions, and it shows clear λ-anomalies at the superconducting transitions. The heat capacity jumps over temperature ($\Delta C_p/T_c$) at the superconducting transitions are estimated using the dashed lines in Fig. 5b. The results give $\Delta C_p/(\gamma T_c)$ = 1.31, 1.47, and 1.52 for $x$ = 0.5, 0.6, and 0.7, respectively, which are close to the expected value of 1.43 for BCS superconductivity in the weak coupling limit.

The fitting of normal state heat capacity data also yields $\beta$ = 3.7(2), 3.8(9), and 4.0(2) mJ/mole/K$^4$ for $x$ = 0.5, 0.6, and 0.7, respectively, and Debye temperatures ($\Theta_D$) are calculated to be 154.0 K, 151.8 K, and 150.1 K using this relation: $\Theta_D = \left(\frac{12\pi^4 nR}{5\beta}\right)^{1/3}$. We found that the crystals with higher In doping concentrations become harder to bend (qualitatively). The electron-phonon coupling constant, $\lambda_{ep}$, can be obtained using the inverted McMillan equation:

$$\lambda_{ep} = \frac{1.04 + \mu^* \ln\left(\frac{\Theta_D}{1.45 T_c}\right)}{(1-0.62\mu^*)\ln\left(\frac{\Theta_D}{1.45 T_c}\right) - 1.04}$$

, where the $\mu^*$ is the Coulomb pseudo potential parameter and is typically given a value of 0.13[35]. The calculation yields $\lambda_{ep}$ = 0.56, 0.59, and 0.60 for $x$ = 0.5, 0.6, and 0.7, respectively. These values all suggest that Pb$_{1-x}$In$_x$Bi$_2$Te$_4$ is a weak-coupling superconductor. The density of states at the Fermi energy $N(E_F)$ can be calculated from $\lambda_{ep}$ and γ by the following equation: $N(E_F) = \frac{3\gamma}{\pi^2 k_B^2 (1+\lambda_{ep})}$, where $k_B$ is the Boltzmann constant. $N(E_F)$ are calculated to be 1.01, 1.11, and 1.12 states $eV^{-1}$ per formula unit for $x$ = 0.5, 0.6, and 0.7, respectively, in Pb$_{1-x}$In$_x$Bi$_2$Te$_4$.

Considering the observed disorder from Indium doping in SC-XRD, an important question is, in the In-doped crystals, how the disorder affects the topology in the band structure. First, though In-Pb-Bi anti-site disorder was observed, the whole chemical stoichiometry does not show significant deviation with nominal based on the EDS and SC-XRD refinement results, which suggests that the total charge in each unit cell and the fermi level should be mostly unaffected by the anti-site disorder. However, the anti-site disorder level is expected to show detectable effect on the band gap and the band topology [36]. Also, note that, the reported Sn and Bi in SnBi$_2$Te$_4$ are mostly ordered [23], which suggests that the anti-site disorder level in In-doped SnBi$_2$Te$_4$ might be lower as well. Consistently, the superconducting transitions of In-doped PbBi$_2$Te$_4$ in Fig. 5 seem less sharp than the reported heat capacity jump in In-doped SnBi$_2$Te$_4$ [23]. Future systematic ARPES measurements in combination with the DFT calculations on In-doped PbBi$_2$Te$_4$ will help to understand the role that anti-site disorder plays in the band structure and potential topological superconductivity.

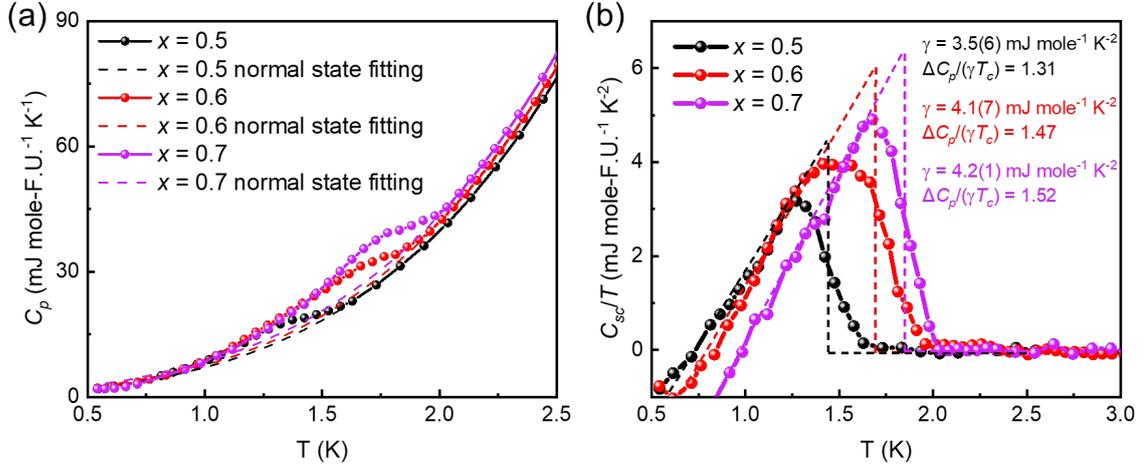

**Figure 5.** (a) The heat capacity as a function of temperature. The dashed lines display the fitting of normal state heat capacity. (b) The contribution of superconducting electrons in heat capacity over temperature. The dashed lines are guides for the eye for the equal area construction used for the estimation of heat capacity jump.

**Conclusions**

In summary, we have experimentally demonstrated type-II bulk superconductivity in $Pb_{1-x}In_xBi_2Te_4$ crystals. In contrast with In-doped $SnBi_2Te_4$ which was found to be *p*-type, the Hall results confirm dominate *n*-type carriers in $Pb_{1-x}In_xBi_2Te_4$. Although these two systems show opposite types of carriers, their superconducting parameters are close to each other. The quantitative analysis of the electron and superconductivity parameters reveals BCS superconductivity in the weak coupling limit. The maximum $T_c$ and $H_{c2}$ (0) are 2.06 K and 1.55 T, achieved for the nominal composition $Pb_{0.3}In_{0.7}Bi_2Te_4$ - both slightly enhanced compared to In-doped $SnBi_2Te_4$. Our findings establish a new ambient-pressure superconducting system, as well as a promising candidate for topological superconductivity. In future investigations, high-resolution electron microscopy or neutron diffraction studies are desired in order to determine the chemical distribution of different atoms in the lattice and understand how they interplay with superconductivity and topology. Our sizable $Pb_{1-x}In_xBi_2Te_4$ single crystals also provide excellent opportunities for future spectroscopic studies to search for topological surface states.

The superconducting $T_c$ monotonically increases with higher In doping concentration although 70% In for Pb substitution seems to be the upper limit of In solubility for ambient pressure synthesis via a conventional solid-state reaction method, a higher In doping concentration or even $InBi_2Te_4$ without changing the structure may be achievable via techniques such as high-pressure synthesis, and there might be a chance for a further promoted superconducting $T_c$.

**Acknowledgements**
The work at Princeton University was supported by the US Department of Energy, Division of Basic Energy Sciences, grant number DE-FG02-98ER45706. W. X. was supported by the US Department of Energy, Division of Basic Energy Sciences, grant number DE-SC0023648.

**References**

[1] X.-L. Qi and S.-C. Zhang, Topological insulators and superconductors, Rev. Mod. Phys. **83**, 1057 (2011).

[2] L. Fu and C. L. Kane, Superconducting Proximity Effect and Majorana Fermions at the Surface of a Topological Insulator, Phys. Rev. Lett. **100**, 096407 (2008).

[3] M. Leijnse and K. Flensberg, Introduction to topological superconductivity and Majorana fermions, Semicond. Sci. Technol. **27**, 124003 (2012).

[4] C. Nayak, S. H. Simon, A. Stern, M. Freedman, and S. Das Sarma, Non-Abelian anyons and topological quantum computation, Rev. Mod. Phys. **80**, 1083 (2008).

[5] S. R. Elliott and M. Franz, Colloquium: Majorana fermions in nuclear, particle, and solid-state physics, Rev. Mod. Phys. **87**, 137 (2015).

[6] D. A. Ivanov, Non-Abelian Statistics of Half-Quantum Vortices in $\mathit{p}$-Wave Superconductors, Phys. Rev. Lett. **86**, 268 (2001).

[7] S. Frolov, M. Manfra, and J. Sau, Topological superconductivity in hybrid devices, Nat. Phys. **16**, 718 (2020).

[8] L. Fu and E. Berg, Odd-Parity Topological Superconductors: Theory and Application to $CuxBi_2Se_3$, Phys. Rev. Lett. **105**, 097001 (2010).

[9] P. Zhang, K. Yaji, T. Hashimoto, Y. Ota, T. Kondo, K. Okazaki, Z. Wang, J. Wen, G. D. Gu, H. Ding, and S. Shin, Observation of topological superconductivity on the surface of an iron-based superconductor, Science **360**, 182 (2018).

[10] G. Xu, B. Lian, P. Tang, X.-L. Qi, and S.-C. Zhang, Topological Superconductivity on the Surface of Fe-Based Superconductors, Phys. Rev. Lett. **117**, 047001 (2016).

[11] S. Sasaki, M. Kriener, K. Segawa, K. Yada, Y. Tanaka, M. Sato, and Y. Ando, Topological Superconductivity in $CuxBi_2Se_3$, Phys. Rev. Lett. **107**, 217001 (2011).

[12] Y. L. Chen, J. G. Analytis, J.-H. Chu, Z. K. Liu, S.-K. Mo, X. L. Qi, H. J. Zhang, D. H. Lu, X. Dai, Z. Fang, S. C. Zhang, I. R. Fisher, Z. Hussain, and Z.-X. Shen, Experimental Realization of a Three-Dimensional Topological Insulator, $Bi_2Te_3$, Science **325**, 178 (2009).

[13] D. Hsieh, Y. Xia, D. Qian, L. Wray, F. Meier, J. H. Dil, J. Osterwalder, L. Patthey, A. V. Fedorov, H. Lin, A. Bansil, D. Grauer, Y. S. Hor, R. J. Cava, and M. Z. Hasan, Observation of Time-Reversal-Protected Single-Dirac-Cone Topological-Insulator States in $Bi_2Te_3$ and $Sb_2Te_3$, Phys. Rev. Lett. **103**, 146401 (2009).

[14] Y. Xia, D. Qian, D. Hsieh, L. Wray, A. Pal, H. Lin, A. Bansil, D. Grauer, Y. S. Hor, R. J. Cava, and M. Z. Hasan, Observation of a large-gap topological-insulator class with a single Dirac cone on the surface, Nat. Phys. **5**, 398 (2009).

[15] J. L. Zhang, S. J. Zhang, H. M. Weng, W. Zhang, L. X. Yang, Q. Q. Liu, S. M. Feng, X. C. Wang, R. C. Yu, L. Z. Cao, L. Wang, W. G. Yang, H. Z. Liu, W. Y. Zhao, S. C. Zhang, X. Dai, Z. Fang, and C. Q. Jin, Pressure-induced superconductivity in topological parent compound $Bi_2Te_3$, Proceedings of the National Academy of Sciences **108**, 24 (2011).

[16] P. P. Kong, J. L. Zhang, S. J. Zhang, J. Zhu, Q. Q. Liu, R. C. Yu, Z. Fang, C. Q. Jin, W. G. Yang, X. H. Yu, J. L. Zhu, and Y. S. Zhao, Superconductivity of the topological insulator $Bi_2Se_3$ at high pressure, J. Phys.: Condens. Matter **25**, 362204 (2013).

[17] Y. S. Hor, A. J. Williams, J. G. Checkelsky, P. Roushan, J. Seo, Q. Xu, H. W. Zandbergen, A. Yazdani, N. P. Ong, and R. J. Cava, Superconductivity in $CuxBi_2Se_3$ and its Implications for Pairing

in the Undoped Topological Insulator, Phys. Rev. Lett. **104**, 057001 (2010).
[18] Y. S. Hor, J. G. Checkelsky, D. Qu, N. P. Ong, and R. J. Cava, Superconductivity and non-metallicity induced by doping the topological insulators Bi2Se3 and Bi2Te3, J. Phys. Chem. Solids **72**, 572 (2011).
[19] J. Shen, W.-Y. He, N. F. Q. Yuan, Z. Huang, C.-w. Cho, S. H. Lee, Y. S. Hor, K. T. Law, and R. Lortz, Nematic topological superconducting phase in Nb-doped Bi2Se3, npj Quantum Mater. **2**, 59 (2017).
[20] Z. Liu, X. Yao, J. Shao, M. Zuo, L. Pi, S. Tan, C. Zhang, and Y. Zhang, Superconductivity with topological surface state in Sr x Bi2Se3, Journal of the American Chemical Society **137**, 10512 (2015).
[21] J. Li, Y. Li, S. Du, Z. Wang, B.-L. Gu, S.-C. Zhang, K. He, W. Duan, and Y. Xu, Intrinsic magnetic topological insulators in van der Waals layered MnBi2Te4-family materials, Sci. Adv. **5**, eaaw5685 (2019).
[22] K. He, MnBi2Te4-family intrinsic magnetic topological materials, npj Quantum Mater. **5**, 90 (2020).
[23] M. A. McGuire, H. Zhang, A. F. May, S. Okamoto, R. G. Moore, X. Wang, C. Girod, S. M. Thomas, F. Ronning, and J. Yan, Superconductivity by alloying the topological insulator SnBi2Te4, Phys. Rev. Mater. **7**, 034802 (2023).
[24] K. Kuroda, H. Miyahara, M. Ye, S. Eremeev, Y. M. Koroteev, E. Krasovskii, E. Chulkov, S. Hiramoto, C. Moriyoshi, and Y. Kuroiwa, Experimental verification of PbBi 2 Te 4 as a 3D topological insulator, Phys. Rev. Lett. **108**, 206803 (2012).
[25] R. Matsumoto, Z. Hou, M. Nagao, S. Adachi, H. Hara, H. Tanaka, K. Nakamura, R. Murakami, S. Yamamoto, and H. Takeya, Data-driven exploration of new pressure-induced superconductivity in PbBi2Te4, Science and Technology of Advanced Materials **19**, 909 (2018).
[26] G. S. Bushmarina, I. A. Drabkin, D. V. Mashovets, R. V. Parfeniev, D. V. Shamshur, and M. A. Shachov, Superconducting properties of the SnTe-PbTe system doped with indium, Physica B **169**, 687 (1991).
[27] M. P. Smylie, K. Kobayashi, J. Z. Dans, H. Hebbeker, R. Chapai, W. K. Kwok, and U. Welp, Full superconducting gap in the candidate topological superconductor In1-xPbxTe for x=0.2, Phys. Rev. B **106**, 054503 (2022).
[28] G. M. Sheldrick, Crystal structure refinement with SHELXL, Acta Crystallographica Section C: Structural Chemistry **71**, 3 (2015).
[29] See Supplemental Material at [URL will be inserted by publisher] for SEM images and EDS spectra, pXRD pattern for Pb$_{0.3}$In$_{0.7}$Bi$_2$Te$_4$, magnetic susceptibility, SC-XRD refinement results and reciprocal planes.
[30] R. Chami and T. JC, CONTRIBUTION A L'ETUDE DU TERNAIRE PLOMB-BISMUTH-TELLURE: ETUDE DE LA COUPE PBTE-BI2-TE3,  (1983).
[31] N. Haldolaarachchige, Q. Gibson, W. Xie, M. B. Nielsen, S. Kushwaha, and R. J. Cava, Anomalous composition dependence of the superconductivity in In-doped SnTe, Phys. Rev. B **93**, 024520 (2016).
[32] A. Bali, R. Chetty, A. Sharma, G. Rogl, P. Heinrich, S. Suwas, D. K. Misra, P. Rogl, E. Bauer, and R. C. Mallik, Thermoelectric properties of In and I doped PbTe, J. Appl. Phys. **120** (2016).
[33] A. Bali, H. Wang, G. J. Snyder, and R. C. Mallik, Thermoelectric properties of indium doped PbTe1-ySey alloys, J. Appl. Phys. **116** (2014).

[34] Y. Matsushita, H. Bluhm, T. Geballe, and I. Fisher, Evidence for charge Kondo effect in superconducting Tl-doped PbTe, Phys. Rev. Lett. **94**, 157002 (2005).

[35] W. L. McMillan, Transition Temperature of Strong-Coupled Superconductors, Physical Review **167**, 331 (1968).

[36] C. Hu, S.-W. Lien, E. Feng, S. Mackey, H.-J. Tien, I. I. Mazin, H. Cao, T.-R. Chang, and N. Ni, Tuning magnetism and band topology through antisite defects in Sb-doped MnBi4Te7, Phys. Rev. B **104**, 054422 (2021).

*Supplementary Information*

**Superconductivity in electron doped PbBi$_2$Te$_4$**

Xianghan Xu[1,*], Danrui Ni[1], Weiwei Xie[2], and R.J. Cava[1,*]

*1. Department of Chemistry, Princeton University, Princeton, NJ 08648*
*2. Department of Chemistry, Michigan State University, Lansing, MI 48824*

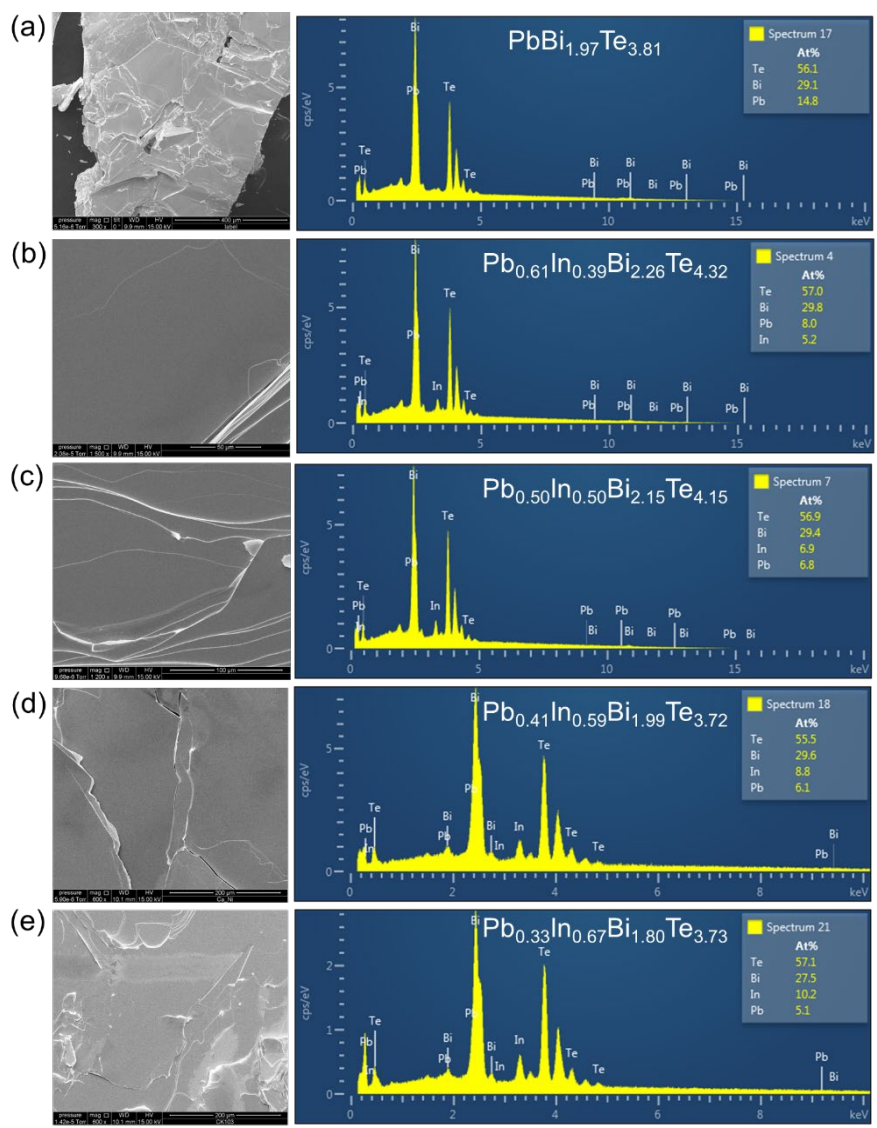

**Figure S1.** (a)-(e) The SEM images and EDS spectra for nominal $x$ = 0, $x$ = 0.4, $x$ = 0.5, $x$ = 0.6, and $x$ = 0.7 Pb$_{1-x}$In$_x$Bi$_2$Te$_4$ crystals.

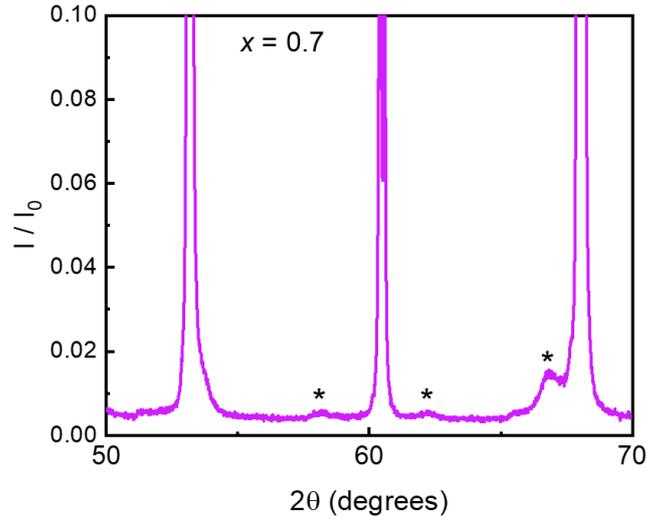

**Figure S2**. Zoomed-in region of the pXRD pattern for a $Pb_{0.3}In_{0.7}Bi_2Te_4$ crystal showing the (00L) reflections with asterisks denoting peaks from trace amount of impurities, which can be indexed by $Bi_2Te_3$. The intensity ($I$) is normalized by the maximum intensity ($I_0$) of the (0 0 30) peak.

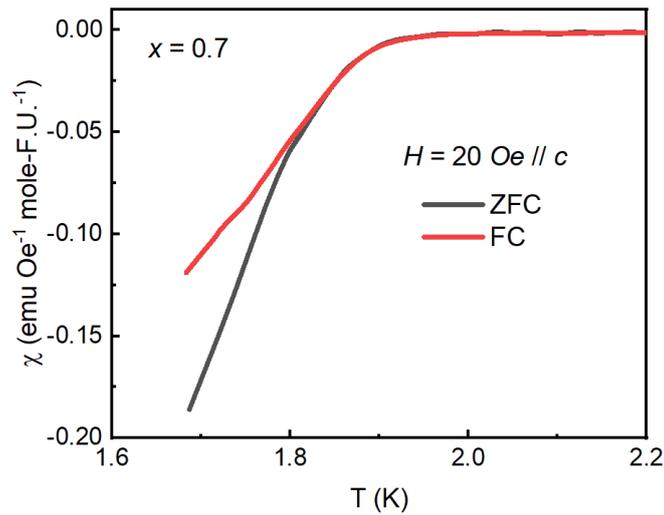

**Figure S3.** The zero-field cooled and field cooled magnetic susceptibility versus temperature on a crystal of $Pb_{0.3}In_{0.7}Bi_2Te_4$ (x = 0.7) measured with a 20 Oe applied magnetic field.

**Table S1**. In occupancies and concentrations from the SC-XRD refinements.

| Nominal composition | In occupancy at Pb sites (%) | In occupancy at Bi sites (%) | Refined In concentration (%) | $R1$, $wR2$, $GooF$ |
|---|---|---|---|---|
| $Pb_{0.6}In_{0.4}Bi_2Te_4$ | 13.68 | 11.54 | 36.76 | 0.0699, 0.1305, 1.178 |
| $Pb_{0.5}In_{0.5}Bi_2Te_4$ | 20.82 | 13.43 | 47.68 | 0.0548, 0.1278, 1.145 |
| $Pb_{0.4}In_{0.6}Bi_2Te_4$ | 15.82 | 19.44 | 54.70 | 0.0557, 0.1262, 1.351 |
| $Pb_{0.3}In_{0.7}Bi_2Te_4$ | 30.91 | 19.67 | 70.25 | 0.0722, 0.1886, 1.274 |

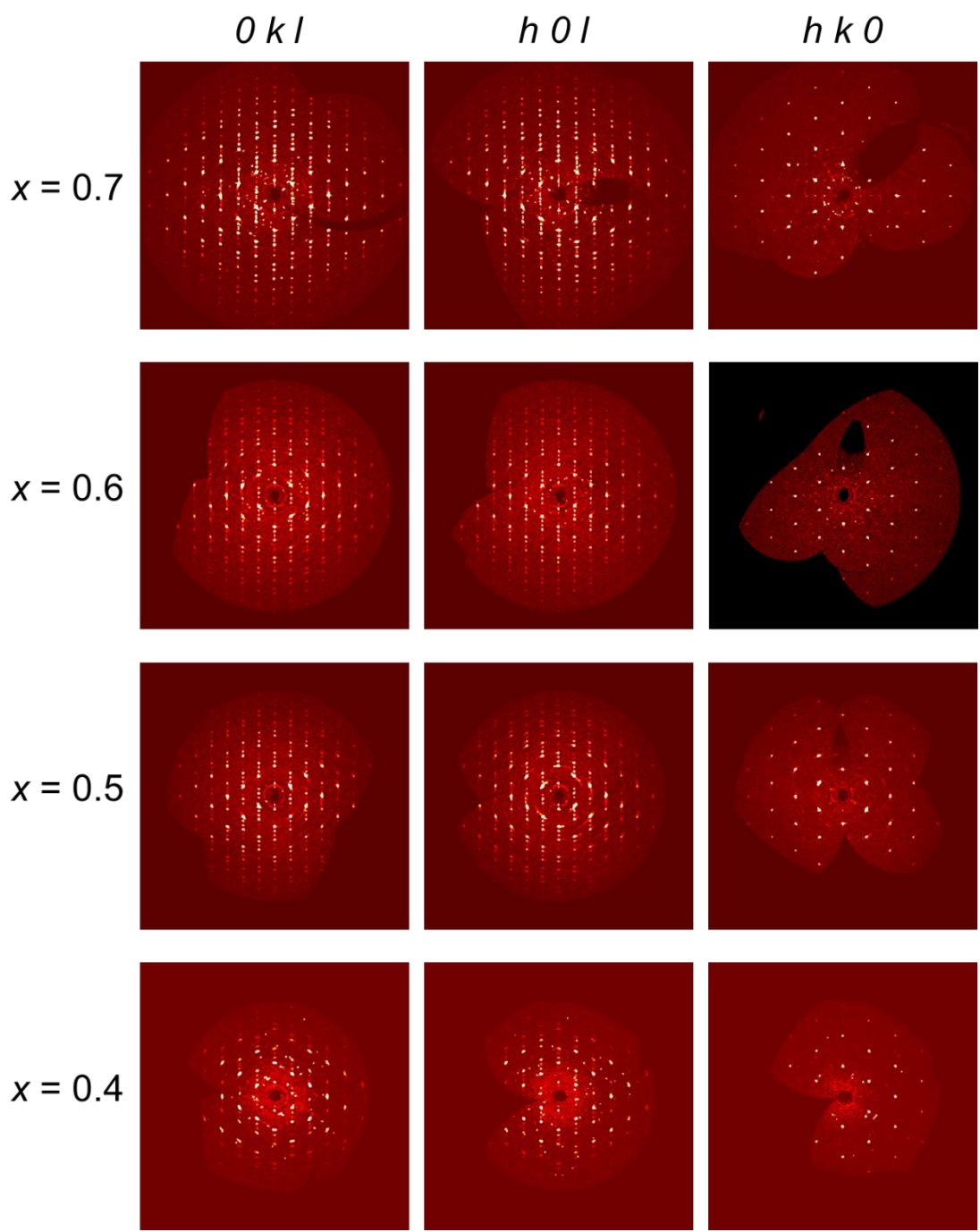

**Figure S4**. (*0 k l*), (*h 0 l*), and (*h k 0*) reciprocal lattice planes of $Pb_{1-x}In_xBi_2Te_4$ crystals obtained from single-crystal x-ray diffractions.

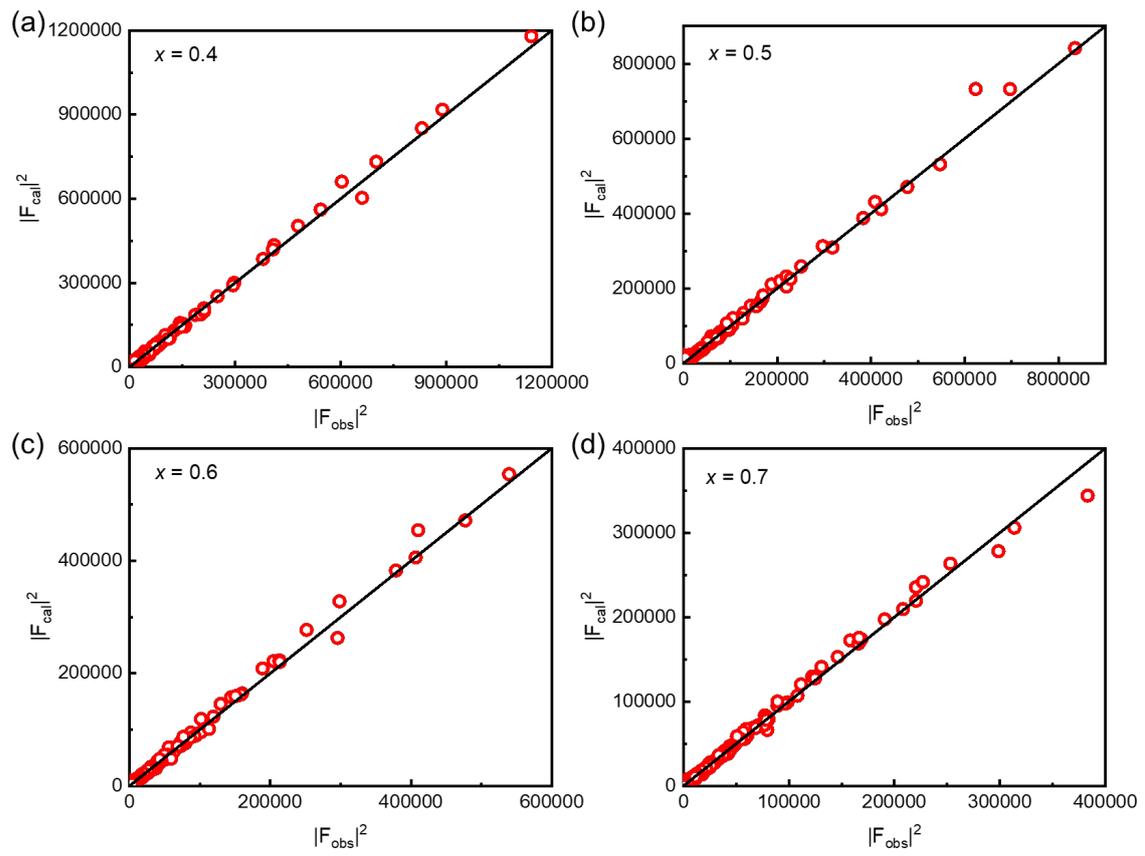

**Figure S5**. $|F_{cal}|^2$ vs. $|F_{obs}|^2$ plots from the refinement results of $Pb_{1-x}In_xBi_2Te_4$ crystals.